\documentclass{article}
\usepackage{graphicx}
\usepackage{color}
\usepackage{amsmath, amsthm}
\usepackage{amssymb}
\usepackage{rotating}
\usepackage{epsfig}
\usepackage{verbatim}
\usepackage{a4}

\input amssym.def
\input amssym.tex

\def\ben{\begin{equation}}
\def\een{\end{equation}}

\def\bea{\begin{eqnarray}}
\def\eea{\end{eqnarray}}

\def\ben{\begin{equation}}
\def\een{\end{equation}}
\def\bea{\begin{eqnarray}}
\def\eea{\end{eqnarray}}

\begin{document}

\title{The Positive Mass and Isoperimetric Inequalities
for Axisymmetric  Black Holes in four and five dimensions}
\author{G. W. Gibbons and G. Holzegel
\\
D.A.M.T.P.,
\\ Cambridge University,
\\ Wilberforce Road,
\\ Cambridge CB3 0WA,
 \\ U.K.}
\date{\today}

\maketitle
\begin{abstract}
In this paper we revisit Brill's proof of positive mass for
three-dimensional, time-symmetric, axisymmetric initial data and 
generalise his argument in various directions. 
In $3+1$ dimensions, we include an apparent horizon in the initial 
data and prove the Riemannian Penrose
inequality in a wide number of cases by an elementary argument. In the
case of $4+1$ dimensions we obtain the analogue of Brill's formula 
for initial data admitting a generalised form of
axisymmetry. Including an apparent horizon in the initial data, 
the Riemannian Penrose inequality is again 
proved for a large class of cases. The results may have applications 
in numerical relativity.
\end{abstract}

%% \vfill \eject
\section{Introduction}
Some  of the most important results in general relativity,
are the various  positive mass theorems for asymptotically
flat and asymptotically Anti-de-Sitter spacetimes (cf. \cite{Schoen,
  Witten, Gibbons, Gibbons4}).
Since the mass in these spacetimes is either constant, or
in the case of the Bondi mass, non-increasing, these theorems
restrict to some degree,  the possible evolution of spacetimes.
Thus, for example since the mass of an  asymptotically flat
spacetime  is bounded below by zero and 
this can only be achieved by flat spacetime, these theorems
are sometimes taken to show that flat spacetime is stable.
However the mass is a single number, determined only
by the asymptotic structure of spacetime. It can give no
detailed information about 
local behaviour of the spacetime geometry. Moreover
one knows from the singularity theorems that under 
some circumstances in which the mass at infinity is positive, 
the occurrence of spacetime singularities 
is inevitable. 

This situation contrasts sharply with that 
for classical field theories in flat spacetime. 
In that case the mass $M$ may be expressed 
as an integral 
\ben 
M= \int _{ {\Bbb  R} ^3} T_{00} \, d^3x \,\, \label{star} 
\een
of a local energy density $T_{00}$ which is typically
positive semi-definite, vanishing only for the trivial
field configuration. For example for a real scalar field $\phi$,

$$
T_{00}= {1 \over 2} \bigl( \nabla \phi \bigr)
 ^2 + {1 \over 2} \bigl 
({\partial \phi \over \partial t} \bigr )^2 + V(\phi)\,, 
$$
where $V(\phi)$ is a positive potential function, with
global minimum at the origin of field space $\phi=0$.

For suitable potential functions $V(\phi)$ the constancy
of the energy places powerful restrictions on how
the field $\phi$ may evolve. In particular, using Sobolev
inequalities  it is possible to show that the evolution
is smooth for all times.

Now because of the Equivalence Principle, there is
no well defined, generally covariant notion, 
of energy density in general relativity, and a
generally covariant  expression like (\ref{star}) is not available.
However, {\sl this does not mean  that one cannot hope to  find
an expression analogous to (\ref{star})} valid
in a particular set of coordinate systems. Moreover, if
the analogue of $T_{00}$ is positive definite
then one may be able to restrict the evolution
of the spacetime, and  perhaps even prove long term
existence. 
  
 Just such a formula was found by Brill \cite{Brill} for
the special case of axisymmetric time-symmetric  initial data
on ${\Bbb  R} ^3$ which are also invariant under reversing 
the azimuthal angle\footnote{that is the metric admits 
an isometric  reversible circle action}.
Using his formula, Brill was able to give the first
mathematically rigorous proof of the positive mass theorem in this 
restricted, but nevertheless physically interesting, special case.  

It is widely appreciated that Brill's result may be extended
in a straightforward way to maximal data. It is possibly less widely
appreciated that the angle reversing assumption can also be
dropped. We will review and generalise Brill's result to this so
called weakly axisymmetric case in section 2.
More interesting physically is the possibility of
extending his theorem to include an apparent horizon
of area $A$ say. In this case one expects, because
of the cosmic censorship hypothesis \cite{Penrose} the stronger 
``isoperimetric'' or ``Riemannian Penrose''
 inequality to hold
\ben
M \ge \sqrt { A \over 16 \pi }\,,\label{starstar}
\een
with equality only for the Schwarzschild initial data. 

In fact (\ref{starstar})  has been proven in the general
time-symmetric case by Huisken and Ilmanen \cite{Huisken} 
using the inverse mean curvature flow (see also \cite{Bray}). 
The purpose of this paper is to show  that Brill's formula may be
extended to incorporate an apparent horizon and,
at least in a large class of cases to be defined precisely below,
to establish the isoperimetric inequality for black holes. These
results are presented in section 3.
     
In recent years, the global behaviour
of higher dimensional spacetimes, 
especially five-dimensional spacetimes, has  
attracted a great deal of interest because of various applications
to String and M-theory, and from  particle physics phenomenologists
interested in scenarios with small extra dimensions 
which might permit the production of small black holes
in accelerators. It is natural, therefore, to enquire
about a possible generalisation of Brill's formula to five dimensions,
that is to four-dimensional initial data sets. 
We will obtain such a generalisation in section 4, subject to the assumption
that the four-dimensional initial data admits  an action
by two commuting circle groups, i.e.~by the torus group $T^2$. 

In five spacetime dimensions, the isoperimetric 
inequality for black holes becomes
$$
M \ge \frac{3\pi}{8} \left(\frac{A}{2 \pi^2}\right)^{\frac{2}{3}} =
\frac{3}{2} \left(\frac{A}{16 \sqrt{\pi}}\right)^{\frac{2}{3}}
$$

As yet, there is not a lot of mathematical evidence
for this inequality. The inverse-mean curvature
method is not available because it makes essential use
of the two-dimensional Gauss-Bonnet theorem.  However
some support is given by the behaviour of collapsing shells 
\cite{Gibbons2}. In section 5 we will use our five-dimensional 
Brill formula to prove the inequality in a wide number of cases.  
\\
\\
Finally, we comment on possible applications of our findings 
to numerical relativity. Three-dimensional axisymmetric 
initial data sets containing black holes have been studied 
numerically for quite some time now (see e.g.~\cite{Bernstein}, where
the superposition of Brill-waves and a black hole is studied). 
Our results concerning four-dimensional initial data, in particular 
formula (\ref{Brilleq}), may be useful for 
controlling the accuracy of numerical codes 
within the study of higher-dimensional gravity.

\section{Brill's proof of positive mass}
Consider a maximal axisymmetric initial data set $(\mathbb{R}^3,
g_{ab},K_{ab})$ for Einstein's equations. This means in particular
that $g$ and $K$ satisfy the constraint equations for $tr K=0$
\begin{align}
R &= 16 \pi \mu + K_{ab}K^{ab} \, , \\
D^a K_{ab} &= 0 \, .
\end{align}
Here $D$ is the covariant derivative with respect to the metric $g$
and $\mu$ is the $T_{00}$ component of the energy momentum
tensor. Axisymmetry means that the metric can be written in the form
\begin{equation} \label{axmet1}
g = q_{AB} dx^A dx^B + X^2 \left(d\phi + A_B dx^B\right)^2 \, ,
\end{equation}
with $q_{AB}$ a two dimensional metric on the orbit space of the
Killing field $\frac{\partial}{\partial \phi}$ and the functions 
$X$ and $A_B$ not depending on the variable $\phi$. If the metric is
\emph{strongly} axisymmetric it enjoys an additional mirror symmetry
and $A$ has to vanish. Coordinates can be found such that
\begin{equation} \label{axmet2}
g = e^{(-2U+2q)} \left(d\rho^2 + dz^2 \right) + \rho^2 e^{-2U}
\left(d\phi + A_{\rho} d\rho + A_z dz \right)^2 \, .
\end{equation}
This choice of coordinates corresponds to finding a harmonic function
on the space of orbits, i.e.~a solution to the equation
\begin{equation}
\bigtriangleup_{q_{AB}} \rho = 0 \, .
\end{equation}
The solution $\rho$ is unique once we specify conditions at infinity and
the $z$-axis, the boundary of the orbit space. Furthermore, to obtain
a regular axisymmetric metric, the function $q$ has to satisfy 
\begin{equation} \label{axcond}
q = 0 \qquad \textrm{and} \qquad \frac{\partial q}{\partial \rho}=0
\qquad \textrm{\ \ on the  z-axis.}
\end{equation}
The first condition is necessary to avoid conical singularities on the
axis. Furthermore we impose that at infinity
\begin{equation} \label{qbehav}
q \sim \frac{1}{r^{1+\epsilon}} \textrm{\ \ \ \ and \ \ \ \ }
\frac{\partial}{\partial r} q \sim
\frac{1}{r^{2+\epsilon}} \, ,
\end{equation}
for some $\epsilon>0$. On the other hand, the $\frac{1}{r}$-term 
of $U$ contains the ADM-mass of the metric:
\begin{equation} \label{ubehav}
U \sim -\frac{m}{r}  \textrm{\ \ \ \ and \ \ \ \ } 
\frac{\partial}{\partial r} U \sim \frac{m}{r^2}
\end{equation}
at infinity.

Let us compute the scalar-curvature of the metric (\ref{axmet2}). The
dreibein is
\begin{equation}
e^1 = e^{-U+q} d\rho \textrm{ \ \ \ , \ \ \ } e^2 = e^{-U+q} dz 
\textrm{ \ \ \ , \ \ \ } e^3 = \rho e^{-U} \left(d\phi + A_{\rho}
d\rho + A_z dz\right) \, .
\end{equation}
The antisymmetric connection coefficients are
\begin{eqnarray}
\omega^1_{\phantom{1}2} &=& \left(\frac{\left(-U+q\right)_{,z}}{e^{-U+q}} e^1 -
\frac{\left(-U+q\right)_{,\rho}}{e^{-U+q}} e^2 \right) + \frac{1}{2}
\rho e^{U-2q} \left(A_{\rho, z} - A_{z,\rho}\right) e^3 \, , \\
\omega^3_{\phantom{3}1} &=& \left(\frac{1}{\rho} - U_{,\rho} \right)
e^{U-q} e^3 - \frac{1}{2} \rho e^{U-2q} \left(A_{\rho, z} -
A_{z,\rho}\right) e^2 \, , \\
\omega^3_{\phantom{3}2} &=& -U_{,z} e^{U-q} e^3 + \frac{1}{2} \rho e^{U-2q} \left(A_{\rho, z} -
A_{z,\rho}\right) e^1 \, .
\end{eqnarray}
Note that indices are raised and lowered with the Euclidean metric. 
The curvature two-form is obtained from
\begin{equation}
R^i_{\phantom{i}j} = d\omega^i_{\phantom{i}j} +
\omega^i_{\phantom{i}k} \wedge \omega^k_{\phantom{k}j} \, .
\end{equation}
Since the scalar curvature is obtained from
\begin{equation} \label{calc}
R = g^{ab} R_{ab} = g^{ab} \left(R^1_{\phantom{1}a1b} + R^2_{\phantom{2}a2b}
+ R^3_{\phantom{3}a3b} \right) = 2R_{1212} + 2R_{1313} + 2R_{2323} \, ,
\end{equation}
we only need the $e^1 \wedge e^2$-part of $R^1_{\phantom{1}2}$, the
$e^1 \wedge e^3$-part of $R^1_{\phantom{1}3}$, and the $e^2 \wedge
e^3$-part of $R^2_{\phantom{2}3}$ to compute the scalar curvature. We obtain
\begin{eqnarray}
R_{1212} &=& e^{2U-2q} \left(\left(U-q\right)_{,\rho \rho} +
\left(U-q\right)_{,zz} \right) - \frac{3}{4} \rho^2 e^{2U-4q}
\left(A_{\rho, z} - A_{z,\rho}\right)^2 \, ,
\end{eqnarray}
\begin{equation}
\begin{split}
R_{1313} = &-\left(\frac{1}{\rho} - U_{,\rho} \right)^2 e^{2U-2q}
-\left(\frac{1}{\rho}e^{U-q} - U_{,\rho} e^{U-q} \right)_{,\rho}
e^{U-q} \\
&+ \left(-U+q\right)_{,z} U_{,z} e^{2U-2q} + \frac{1}{4} \rho^2 e^{2U-4q}
\left(A_{\rho, z} - A_{z,\rho}\right)^2 \, ,
\end{split}
\end{equation}
\begin{equation}
\begin{split}
R_{2323} = &-\left(U_{,z}\right)^2 e^{2U-2q} + \left(e^{U-q} U_{,z}
\right)_{,z} e^{U-q} \\
&+ \frac{1}{4} \rho^2 e^{2U-4q} \left(A_{\rho, z} -
A_{z,\rho}\right)^2 + \left(U-q\right)_{,\rho} \left(\frac{1}{\rho} -
U_{,\rho} \right) e^{2U-2q} \, .
\end{split}
\end{equation}
From (\ref{calc}) the scalar curvature then is
\begin{equation}
\begin{split}
\phantom{}^{(3)}R = 2 e^{2U-2q} \left(\left(U-q\right)_{,\rho \rho} +
\left(U-q\right)_{,z z} \right) - \frac{1}{2} \rho^2 e^{2U-4q}
\left(A_{\rho, z} - A_{z,\rho}\right)^2 \\ +  
e^{2U-2q} 2 \left(U_{,\rho \rho} + U_{,z z} \right) 
- 2 e^{2U-2q} \left(\left(U_{,\rho}\right)^2 + \left(U_{,z}\right)^2
\right) + e^{2U-2q} \frac{4}{\rho} U_{,\rho} \, ,
\end{split}
\end{equation} 
which we can write in the form
\begin{equation} \label{diffform}
-\frac{1}{8} \phantom{}^{(3)}R \cdot e^{(-2U+2q)} = -\frac{1}{2}
 \bigtriangleup U + \frac{1}{4} \left( \nabla U \right)^2 +
 \frac{1}{4} \bigtriangleup q - \frac{1}{4}\frac{1}{\rho}\frac{\partial
 q}{\partial \rho} + \frac{1}{16} \rho^2 e^{-2q} \left(A_{\rho, z} -
 A_{z,\rho}\right)^2 \, .
\end{equation}
The Laplacian and gradient are taken with respect to flat
coordinates in $\mathbb{R}^3$. We can integrate this expression over
$\mathbb{R}^3$ to prove positive mass. If we take conditions
(\ref{axcond}),(\ref{qbehav}),(\ref{ubehav}) into account the
integration yields
\begin{equation} \label{mf}
M = \frac{1}{16 \pi} \int \Big[\phantom{}^{(3)}R + \frac{1}{2} \rho^2
  e^{-4q+2U}\left(A_{\rho, z} - A_{z,\rho}\right)^2 \Big] e^{2q}e^{-2U} d^3 x +
\frac{1}{8\pi}\int \left(\nabla U\right)^2 d^3x \, .
\end{equation}
Defining
\begin{equation}
H_{AB} = A_{A,B} - A_{B,A} \, ,
\end{equation}
where $A,B \in \{ 1,2 \}$ denote indices on the orbit space with metric
$q_{AB}$ defined above, and noting that $X=\rho e^{-U}$, we can rewrite
(\ref{mf}) in the neat form 
\begin{equation}
M = \frac{1}{16 \pi} \int \Big[\phantom{}^{(3)}R + \frac{1}{4} X^2
  H_{AB}H^{AB} \Big] \cdot e^{2q}e^{-2U} d^3 x +
\frac{1}{8\pi}\int \left(\nabla U\right)^2 d^3x \, .
\end{equation}
Since $\phantom{}^{(3)}R=16 \pi \mu + K_{ab}K^{ab} \geq 0$ this proves positive mass
for the type of initial data considered above.

\section{The Generalisation to Black Holes}
In this section we will generalise Brill's formula to 
initial data leading to black holes. To determine the 
event horizon of a black hole in a four-dimensional 
spacetime it is necessary to know its entire global 
structure, i.e.~there is no direct way to see an 
event horizon in the initial data. 
On the other hand, if we consider initial data whose maximal
development admits a complete null-infinity, an apparent horizon 
present in the initial data will always lead to a regular event horizon 
in the development (cf. \cite{Hawking}). 
Moreover, in this case there will 
exist on the initial data slice a component of the event horizon 
outside the apparent horizon or coinciding with it. 
We note that for time-symmetric and also for $t-\phi$-symmetric 
initial-data\footnote{A $t-\phi$--symmetric initial data set is 
defined to be an initial data set leading to a $t-\phi$-symmetric 
spacetime, i.e.~a spacetime symmetric under a simultaneous change of
sign of the time coordinate $t$ and the azimuthal coordinate $\phi$.}
 the notion of an apparent horizon and a minimal 
surface coincide. We will restrict to strongly axisymmetric data in
the following.

\subsection{Static initial data: Weyl's class}
Consider Weyl's family of axisymmetric static vacuum solutions to
Einstein's equations:
\begin{equation} \label{Weyl}
ds^2 = e^{2U\left(\rho,z\right)}dt^2 -
e^{-2U\left(\rho,z\right)}\left[e^{2q\left(\rho,z\right)}\left(d\rho^2
+ dz^2\right) + \rho^2d\phi^2\right] \, ,
\end{equation}
with the functions $U$ and $q$ satisfying the equations
\begin{eqnarray} \label{weyleq}
\bigtriangleup U &=& 0 \, , \nonumber \\
\partial_\rho q &=& \rho\left[\left(\partial_\rho U\right)^2 -
  \left(\partial_z U\right)^2\right] \, ,  \\ \nonumber
\partial_z q &=& 2\rho \partial_\rho U \partial_z U \, .
\end{eqnarray}
Note that the natural spatial slices of (\ref{Weyl}) are 
precisely the ones under consideration (\ref{axmet2}) as initial data slices. 
If the topology of the initial data slice is trivial,
i.e.~$\mathbb{R}^3$, then by Liouville's theorem, $U=0$ ($U$ vanishes
 at infinity due to asymptotic flatness). The only static solution 
 in Weyl's class with $\mathbb{R}^3$ topology is Minkowski
 space. To get non-trivial static initial data, we will need
 to allow non-trivial topology. The standard example is provided by 
the Schwarzschild data discussed in the next subsection. Prior to this 
we explicitly check the validity of Brill's equation (\ref{diffform}) 
for the spatial slices in Weyl's class. Using the 
equations (\ref{weyleq}) one easily shows that
\begin{equation}
\frac{1}{4}\left(\nabla U\right)^2 + \frac{1}{4} \bigtriangleup q -
\frac{1}{4}\frac{1}{\rho}\frac{\partial q}{\partial \rho} = 0 \, .
\end{equation} 
Since also  $\bigtriangleup U=0$ and $R=0$ 
for the spatial slices in Weyl's class, equation (\ref{diffform}) holds.

\subsection{Schwarzschild initial data} 
The function $U$ in (\ref{Weyl}) leading to Schwarzschild data can be
obtained by calculating the Newtonian potential of a rod with unit
mass density occupying the z-axis from $-M$ to $M$, as depicted in the
figure below. 
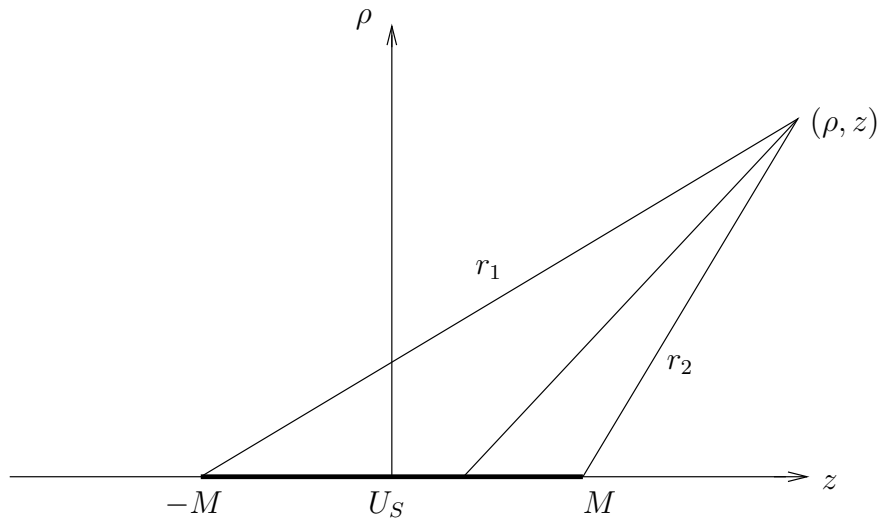
\[
\begin{picture}(0,0)%
\includegraphics{Weyl2.pstex}%
\end{picture}%
\setlength{\unitlength}{3947sp}%
\begingroup\makeatletter\ifx\SetFigFont\undefined%
\gdef\SetFigFont#1#2#3#4#5{%
  \reset@font\fontsize{#1}{#2pt}%
  \fontfamily{#3}\fontseries{#4}\fontshape{#5}%
  \selectfont}%
\fi\endgroup%
\begin{picture}(5388,3265)(1189,-3044)
\put(6301,-2836){\makebox(0,0)[lb]{\smash{{\SetFigFont{12}{14.4}{\rmdefault}{\mddefault}{\updefault}{\color[rgb]{0,0,0}$z$}%
}}}}
\put(4126,-1486){\makebox(0,0)[lb]{\smash{{\SetFigFont{12}{14.4}{\rmdefault}{\mddefault}{\updefault}{\color[rgb]{0,0,0}$r_1$}%
}}}}
\put(5326,-2086){\makebox(0,0)[lb]{\smash{{\SetFigFont{12}{14.4}{\rmdefault}{\mddefault}{\updefault}{\color[rgb]{0,0,0}$r_2$}%
}}}}
\put(6226,-586){\makebox(0,0)[lb]{\smash{{\SetFigFont{12}{14.4}{\rmdefault}{\mddefault}{\updefault}{\color[rgb]{0,0,0}$(\rho,z)$}%
}}}}
\put(3376, 89){\makebox(0,0)[lb]{\smash{{\SetFigFont{12}{14.4}{\rmdefault}{\mddefault}{\updefault}{\color[rgb]{0,0,0}$\rho$}%
}}}}
\put(2176,-2986){\makebox(0,0)[lb]{\smash{{\SetFigFont{12}{14.4}{\rmdefault}{\mddefault}{\updefault}{\color[rgb]{0,0,0}$-M$}%
}}}}
\put(4801,-2986){\makebox(0,0)[lb]{\smash{{\SetFigFont{12}{14.4}{\rmdefault}{\mddefault}{\updefault}{\color[rgb]{0,0,0}$M$}%
}}}}
\put(3451,-2986){\makebox(0,0)[lb]{\smash{{\SetFigFont{12}{14.4}{\rmdefault}{\mddefault}{\updefault}{\color[rgb]{0,0,0}$U_S$}%
}}}}
\end{picture}%

\]
We note the relations
\begin{eqnarray}
r_1^2 &=& \rho^2 + \left(z+M\right)^2 \, , \\
r_2^2 &=& \rho^2 + \left(z-M\right)^2 \, .
\end{eqnarray}
The Newtonian potential at the point $(\rho,z)$ is
\begin{equation} \label{ueq}
\begin{split}
U_S\left(\rho,z\right) &= -\frac{1}{2} \int_{-M}^M
\frac{1}{\sqrt{\rho^2+\left(z-\tilde{z}\right)^2}} d\tilde{z} \\ &= -\frac{1}{2}
\log\left(\frac{M+z+r_1}{-M+z+r_2}\right) =
-\frac{1}{2}\log\left(\frac{r_1+r_2+2M}{r_1+r_2-2M}\right) \, .
\end{split}
\end{equation}
The function $q$ is then found from the equations (\ref{weyleq}) to be 
\begin{equation} \label{qeq}
q_S\left(\rho,z \right) = \frac{1}{2} \log
\left(\frac{\left(r_1+r_2\right)^2 - 4M^2}{4r_1r_2}\right) \, .
\end{equation}
It will be useful to work in spheroidal coordinates $(\lambda,
\mu)$. They are defined by the relations
\begin{eqnarray}
2\lambda &=& r_1+r_2 \, ,\\
2 \mu M &=& r_1-r_2 \, .
\end{eqnarray}
with inverse
\begin{eqnarray}
r_1 &=& \lambda + \mu M \, , \\
r_2 &=& \lambda - \mu M \, ,
\end{eqnarray}
and hence
\begin{eqnarray} \label{sphero}
\rho^2 &=& \left(\lambda^2-M^2\right)\left(1-\mu^2\right) \, , \\
z &=& \lambda \mu \, .
\end{eqnarray}
Note that $\mu \in \left[-1,1\right]$. The transformations
\begin{equation}
\frac{\partial}{\partial \lambda} = \mu \frac{\partial}{\partial z} +
\lambda \sqrt{\frac{1-\mu^2}{\lambda^2-M^2}} \frac{\partial}{\partial
  \rho} \textrm{ \ \ \ \ and \ \ \ \ } \frac{\partial}{\partial \mu} = \lambda \frac{\partial}{\partial z} -
\mu \sqrt{\frac{\lambda^2-M^2}{1-\mu^2}} \frac{\partial}{\partial
  \rho} \, ,
\end{equation}
and hence
\begin{equation} \label{derh1}
\frac{\partial}{\partial \rho} =
\frac{\sqrt{\left(\lambda^2-M^2\right)\left(1-\mu^2\right)}}{\lambda^2-\mu^2M^2}
\left(\lambda \partial_\lambda - \mu \partial_\mu \right) \, ,
\end{equation}
\begin{equation} \label{derh2}
\frac{\partial}{\partial z} =\frac{1}{\lambda^2-\mu^2 M^2} \Big( \mu \left(\lambda^2-M^2\right)
\partial_\lambda + \lambda\left(1-\mu^2\right) \partial_\mu \Big) \, ,
\end{equation}
will turn out to be useful later. In spheroidal coordinates we have
\begin{equation}
U_S = -\frac{1}{2} \log \left(\frac{\lambda+M}{\lambda-M}\right) \, ,
\end{equation}
\begin{equation}
q_S = \frac{1}{2} \log
\left(\frac{\lambda^2-M^2}{\lambda^2-\mu^2M^2}\right) \, .
\end{equation}
One finds that
\begin{equation}
d\rho^2 + dz^2 + \rho^2 d\phi^2 = \left(\lambda^2-\mu^2M^2\right)
\left(\frac{d\lambda^2}{\lambda^2-M^2} + \frac{d\mu^2}{1-\mu^2}\right)
+ \left(\lambda^2-M^2\right)\left(1-\mu^2\right) d\phi^2 \, ,
\end{equation}
and hence the metric (\ref{Weyl}) becomes
\begin{equation}
ds^2 = \frac{\lambda-M}{\lambda+M} dt^2 - \frac{\lambda+M}{\lambda-M}
d\lambda^2 - \frac{\left(\lambda+M\right)^2}{1-\mu^2} d\mu^2 -
\left(\left(\lambda+M\right)^2 \left(1-\mu^2\right) \right) d\phi^2 \, .
\end{equation}
To see that this is in fact the Schwarzschild metric, we finally
transform to coordinates
\begin{eqnarray}
\lambda = r-M \, , \\
\mu = \cos \theta \, ,
\end{eqnarray}
to find 
\begin{equation}
ds^2 = \left(1-\frac{2M}{r}\right)dt^2 -
\left(1-\frac{2M}{r}\right)^{-1} dr^2 - r^2 \left(d\theta^2 +\sin^2
\theta d\phi^2\right) \, .
\end{equation}
{\bf Summary:} We found that an initial data slice for the Schwarzschild
metric can be retrieved from the ansatz (\ref{axmet2}) 
for arbitrary axisymmetric initial data by choosing $U$ and $q$ 
as given in (\ref{ueq}) and (\ref{qeq}). \\ \\
{\bf Remark:} This is just one possible choice. Another example is
provided by choosing $q=0$ and
$e^{-2U}=\left(1+\frac{M}{2r}\right)^4$, which leads
to a conformally flat, time-symmetric Kruskal slice in the
Schwarzschild spacetime. However, these spatial slices are not the
ones used in the Weyl ansatz (\ref{Weyl}). \\ \\
We see that that in the Weyl-form of the metric the 
Schwarzschild horizon is represented by a 
rod in the initial data. On the rod $\lambda=M$ and 
$\mu$ runs from $-1$ to $1$. The rod has area
\begin{equation}
A = \int_0^{2\pi} \int_{-M}^M e^{-2U_S+q_S} \rho dz d\phi =
\int_0^{2\pi} \int_{-1}^1 4M^2 d\mu d\phi = 16 \pi M^2
\end{equation}
and the normal derivative 
\begin{equation} \label{normder}
e^{U_S-q_S}\frac{\partial}{\partial \rho} A \Bigg|_{rod} = \frac{1}{2}
\sqrt{1-\mu^2} \int_0^{2\pi} \int_{-M}^M
e^{-2U_S+q_S} 
\left(1+ \rho \partial_\rho \left(-2U_S + q_S\right)\right) dz d\phi = 0
\end{equation}
vanishes on the rod showing that it is indeed a minimal surface. 
Equation (\ref{normder}) is most easily seen in 
spheroidal coordinates, where one computes (now using (\ref{derh1}) and (\ref{derh2})) 
\begin{equation}
\rho \frac{\partial}{\partial \rho} U_S = M \lambda
\frac{1-\mu^2}{\lambda^2-\mu^2M^2} \, ,
\end{equation}
and hence
\begin{equation} \label{Urod}
\rho \frac{\partial}{\partial \rho} U_S \Bigg|_{rod} = 1 \, ,
\end{equation}
as well as
\begin{equation}
\rho \frac{\partial}{\partial \rho} q_S = 
M^2 \frac{1-\mu^2}{\lambda^2-\mu^2M^2} \, ,
\end{equation}
and therefore
\begin{equation} \label{qrod}
\rho \frac{\partial}{\partial \rho} q_S \Bigg|_{rod} = 1 \, ,
\end{equation}
from which (\ref{normder}) follows. 

\begin{comment}
Upon integration we will have to take the rod as an extra 
boundary into account....
\begin{equation}
\int_{rod} \vec{n} \vec{\nabla} U_S = 2\pi \int_{-M}^M \rho \partial_\rho U_S
\Bigg|_{rod} dz = 4\pi M = 
\int_{\infty} \vec{n} \vec{\nabla} U = \lim_{r \rightarrow \infty}
\left(r^2 \frac{\partial}{\partial r} U_S \right) \cdot 4\pi = 4\pi M
\end{equation}
which is obviously true. 
\end{comment}

\subsection{Non-static data: Deformations of Schwarzschild}
Consider now a deformation of the Schwarzschild data.
\begin{eqnarray}
U = U_S + \tilde{U} \textrm{\ \ \ \ and \ \ \ \ } q = q_S + \tilde{q} \, ,
\end{eqnarray}
where we make the following assumptions about the deformation
$(\tilde{U},\tilde{q})$:
\begin{enumerate}
\item No new minimal surface should be created by the
  deformation. (Note however, that the deformation does not have to be
  small in general.)
\item The rod should remain a minimal surface. This implies that
  $\partial_{\rho} \left(2\tilde{U}-\tilde{q}\right)=0$ will hold on the
  rod. 
\item The area of the rod should be unaltered under the deformation: \\
  $A=8 \pi M \int_{-M}^M \exp\left(-2\tilde{U}+\tilde{q}\right)
  dz=16\pi M^2$.
\end{enumerate}
We will eventually need the stronger assumption \\ \\
$\phantom{XI} 3^\prime.$ The deformation $(\tilde{U},\tilde{q})$ has
support only outside the rod, \\
$\phantom{XXiI}$ i.e.~in particular $\tilde{U}=\tilde{q}=0$ on the
rod. \\ \\ 
Assumption $3^\prime$ implies $2$ and $3$. 
Near infinity we assume the asymptotics
\begin{eqnarray}
\tilde{U} \sim \frac{-\Delta M}{r} \textrm{\ \ \ \ and \ \ \ \ }
\tilde{q} \sim \frac{1}{r^{1+\epsilon}} \, ,
\end{eqnarray}
and that derivatives lower the power of $r$ by one. 
Integrating (\ref{diffform}) now yields the formula
\begin{eqnarray}
\Delta M &= \frac{1}{16 \pi} \int_{vol} R e^{-2U+2q} + \frac{1}{8\pi}
\int_{vol} \left(\nabla \tilde{U} \right)^2 - \frac{1}{4 \pi} \int_{rod}
\left(-\vec{n}\right) \vec{\nabla} \left(\tilde{U} - \frac{1}{2}
\tilde{q} \right)
\nonumber \\ &+ \frac{1}{4} \int_{-M}^M dz \tilde{q} + \frac{1}{4 \pi}
\int_{rod} \tilde{U} \left(-\vec{n}\right) \vec{\nabla} U_S - \frac{1}{4\pi}
\int_{vol} \tilde{U} \bigtriangleup U_S \, ,
\end{eqnarray}
where the last term vanishes because $U_S$ is harmonic. \\
If we introduce the quantity
\begin{equation}
E = \frac{1}{16 \pi} \int_{vol} R e^{-2U+2q} + \frac{1}{8\pi}
\int_{vol} \left(\nabla \tilde{U} \right)^2 \geq 0 \, ,
\end{equation}
we can write 
\begin{equation} \label{deltam}
\Delta M = E - \frac{1}{2} \int_{-M}^M \left(\tilde{U} - \frac{1}{2}
\tilde{q}\right)\Bigg|_{\rho=0} dz + \frac{1}{2}  
\int_{-M}^M \rho \frac{\partial}{\partial \rho} \left(\tilde{U} - \frac{1}{2} \tilde{q}\right)\Bigg|_{\rho=0} dz 
\end{equation}
describing the change of mass under the deformation
considered. The last term of (\ref{deltam}) vanishes by assumption $2$.
Unfortunately, assumption $3$ will only give an upper bound on the
second term of (\ref{deltam}), since by assumption 3 and the
inequality $e^x \geq 1+x$ we have
\begin{equation}
A=16\pi M^2 = 8 \pi M \int_{-M}^M \exp\left(-2\tilde{U}+\tilde{q}\right)
  dz \geq 16\pi M^2 + 8\pi M \int_{-M}^M
  \left(-2\tilde{U}+\tilde{q}\right) dz \, ,
\end{equation}
and therefore 
\begin{equation}
-\frac{1}{2} \int_{-M}^M \left(\tilde{U}-\frac{1}{2}\tilde{q}\right)
 dz \leq 0 \, .
\end{equation}
Consequently, the second term in (\ref{deltam}) 
could in principle spoil the inequality we
want to obtain. On the other hand, assuming $3^\prime$, 
equation (\ref{deltam}) immediately yields a version of the
Riemannian Penrose inequality: \\ \\
{\emph{Lemma 3.1:}} For axisymmetric deformations $(\tilde{U},
\tilde{q})$ of Schwarzschild initial-data satisfying assumptions $1$
and $3^\prime$ the Riemannian-Penrose inequality 
\begin{equation} \label{rpie}
M \geq \sqrt{\frac{A}{16\pi}}
\end{equation}
holds.
\begin{proof} 
If the deformation has support only outside the rod, both
integral terms in (\ref{deltam}) vanish. Consequently, the mass
increases by $E$ while the area remains unaltered. 
\end{proof}
One would really like to show more, namely that the inequality
(\ref{rpie}) does not only hold for arbitrary axisymmetric 
deformations of Schwarzschild away from the horizon 
but for general axisymmetric initial data containing a single horizon. 
One would like to show the following statement, referred to in the
future as \emph{``RPI''}.
\\
\\
\emph{Riemann-Penrose-inequality (RPI): } Given any 
asymptotically-flat, axisymmetric initial data 
with apparent horizon\footnote{We remind the reader that for $t-\phi$
  symmetric initial-data the notion of an apparent horizon and a
  minimal surface coincide.} of area $A$. Then (\ref{rpie})
holds with equality if and only if the initial data is Schwarzschild.
\\
\\
Though we will not be able to establish the above statement by our methods in
this paper, we will indicate in the following how one could proceed
and what final obstacle one encounters. 

To establish \emph{RPI} it would suffice to show 
that any axisymmetric initial 
data can be written in Weyl-form as a ``deformed Schwarzschild data'' 
with the deformation satisfying $-2\tilde{U}+\tilde{q}=0$ on the rod.
The following lemma shows that it is at least always 
possible to transform the initial data 
into Weyl form (\ref{weylform}) using a harmonic map. 
However, we cannot prove that the functions $U$ and $q$ obtained
by bringing the data into Weyl-form have to 
agree with their corresponding Schwarzschild values on the rod.
\\
\\
\emph{Lemma 3.2:} Any axisymmetric initial data set with single 
connected apparent horizon of area $A$, 
can -- in some coordinates -- be written in the form
\begin{equation} \label{weylform}
g = e^{-2U+2q} \left(d\rho^2 + dz^2 \right) + \rho^2 e^{-2U} d\phi^2
\, ,
\end{equation}
such that the minimal surface is a rod of area A on the z-axis
between $-M$ and $M$ for some $M$.
\begin{proof}
Given any axisymmetric initial data with 
apparent horizon of area $A$, we can construct harmonic coordinates 
$\left(\hat{\rho},\hat{z}\right)$ by solving a particular conformal 
mapping problem. We have to solve
\begin{equation}
\bigtriangleup \hat{\rho}\left(\rho,z\right) = 0
\end{equation}
with boundary condition $\hat{\rho}=0$ on the $\rho$-axis and on the
apparent horizon $T$ and $\hat{\rho}$ should tend to the distance from
the axis, i.e.~$\rho$, at infinity.
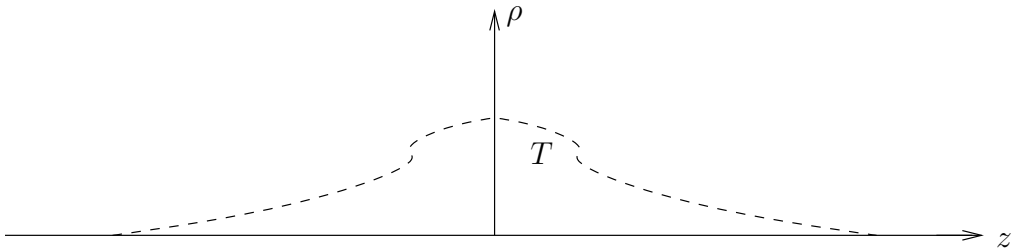
\[
\begin{picture}(0,0)%
\includegraphics{minaxi.pstex}%
\end{picture}%
\setlength{\unitlength}{3947sp}%
\begingroup\makeatletter\ifx\SetFigFont\undefined%
\gdef\SetFigFont#1#2#3#4#5{%
  \reset@font\fontsize{#1}{#2pt}%
  \fontfamily{#3}\fontseries{#4}\fontshape{#5}%
  \selectfont}%
\fi\endgroup%
\begin{picture}(6513,1606)(1114,-2885)
\put(4426,-2311){\makebox(0,0)[lb]{\smash{{\SetFigFont{12}{14.4}{\rmdefault}{\mddefault}{\updefault}{\color[rgb]{0,0,0}$T$}%
}}}}
\put(4276,-1411){\makebox(0,0)[lb]{\smash{{\SetFigFont{12}{14.4}{\rmdefault}{\mddefault}{\updefault}{\color[rgb]{0,0,0}$\rho$}%
}}}}
\put(7351,-2836){\makebox(0,0)[lb]{\smash{{\SetFigFont{12}{14.4}{\rmdefault}{\mddefault}{\updefault}{\color[rgb]{0,0,0}$z$}%
}}}}
\end{picture}%

\]
This is a classical Dirichlet-problem with Lipschitz boundary,
admitting a unique solution. The harmonic analogue of $\hat{\rho}$,
$\hat{z}$, follows from the Cauchy-Riemann equations
\begin{eqnarray}
\partial_{\rho} \hat{z} &=& -\partial_z \hat{\rho} \, , \\
\partial_{z} \hat{z} &=& \partial_\rho \hat{\rho} \, .
\end{eqnarray}
The coordinate transformation $(\rho,z) \rightarrow
(\hat{\rho},\hat{z})$ maps the minimal surface to a rod on the
$z$-axis, lying between $-M$ and $M$, say. In these coordinates, the
metric can be written as
\begin{equation} \label{confmet}
g = e^{(-2U+2q)} \left(d\hat{\rho}^2 + d\hat{z}^2 \right) + \hat{\rho}^2 e^{-2U} d\phi^2
\end{equation}
for some functions $U$ and $q$, unique up to a constant, which is
determined by asymptotic flatness. 
\end{proof}
Now given any axisymmetric initial data with apparent horizon of area
$A$, we can apply {\emph{Lemma 3.2},
mapping the minimal surface to a rod between $-M$ and
$M$, say, and obtaining unique 
functions $U$ and $q$, which are obviously 
singular on the rod. If we split them according to
\begin{equation}
U = U_S + \tilde{U} \textrm{\ \ \ \ and \ \ \ \ } q = q_S + \tilde{q}
\end{equation}
where we choose $U_S$, $q_S$ as in (\ref{ueq}) and (\ref{qeq}), 
we find that assumption 2 is satisfied and 
the second term in (\ref{deltam}) vanishes. Unfortunately we do not
know anything about the behaviour of $\tilde{U}, \tilde{q}$ (and hence
about the first term in (\ref{deltam})) on the rod. One way to prove
{\emph{RPI}} would therefore be to show that the functions $U$ and $q$
obtained implicitly by {\emph{Lemma 3.2}} satisfy the equation
\begin{equation}
-2U+q = -2U_S + q_S
\end{equation}
on the rod. This would mean that the functions $U$ and $q$, 
however complicated they look like globally, 
always admit the same blow-up behaviour as in the
Schwarzschild case, when restricted to the rod. 
However, we have been unable to show that this
follows from the harmonic map of {\emph{Lemma 3.2}} 
and therefore the general case remains
unproved by our methods.

\section{Generalisation to $4+1$ dimensions: Regular data}
In this section we consider regular four-dimensional initial-data for 
the five-dimensional theory of Einstein-gravity. The initial data 
should admit two Killing vectors with $U(1)$-orbits. 

\subsection{Preliminaries}
The general metric has the form
\begin{align} \label{4met}
\begin{split}
g &= q_{AB}dx^A dx^B + Y^2 d\alpha^2 + Z^2 d\alpha d\beta + X^2 d\beta^2  \\
  &= e^{\left(2f+U+V\right)(\rho,\omega)} \left(d\rho^2 +
  \rho^2 d \omega^2 \right) +
  W(\rho,\omega)e^{U(\rho,\omega)+V(\rho,\omega)}\rho^2 \sin \omega 
\cos \omega d\alpha d\beta \\
  & \ \ \  
+ e^{2U(\rho,\omega)}\rho^2
\sin^2 \omega d\alpha^2 + e^{2V(\rho,\omega)}\rho^2 \cos^2 \omega
  d\beta^2 \, .
\end{split}
\end{align}
Without the cross-term $W$ we would have rectangular tori. In general 
$\left|W\right|<1$ must hold to ensure that the metric on the torus is
positive definite. Note that the choice $f=U=V=W=0$ corresponds 
to the flat metric 
\begin{equation}
\delta = dx^2+dy^2+du^2+dv^2
\end{equation}
written in the coordinates
\begin{equation}
x+iy = \rho \sin \omega e^{i \alpha} \qquad u+iv = \rho \cos \omega
e^{i \beta} \, .
\end{equation}
The range of coordinates for (\ref{4met}) is
\begin{equation}
\alpha \in (0, 2\pi] \qquad \beta \in (0,2\pi] \qquad \omega \in
    (0,\frac{\pi}{2}) \, .
\end{equation} 
There are coordinate singularities at $\omega=0$ and
$\omega=\frac{\pi}{2}$. To avoid conical singularities we have to
assume certain behaviour of the functions $f,U,V,W$. We write the
metric in the orthogonal coordinates given above
\begin{align}
\begin{split}
g = dx^2 \left( \frac{e^{2f+U+V}x^2}{x^2+y^2} +
\frac{e^{2U}y^2}{x^2+y^2}\right) + dy^2 \left( \frac{e^{2f+U+V}y^2}{x^2+y^2} +
\frac{e^{2U}x^2}{x^2+y^2}\right) + \\
 2dx \cdot dy \cdot xy \left(
\frac{e^{2f+U+V}}{x^2+y^2} -\frac{e^{2U}}{x^2+y^2}\right) + 2du \cdot dv \cdot uv \left(
\frac{e^{2f+U+V}}{u^2+v^2} -\frac{e^{2V}}{u^2+v^2}\right) \\
+du^2 \left( \frac{e^{2f+U+V}u^2}{u^2+v^2} +
\frac{e^{2V}v^2}{u^2+v^2}\right) + dv^2 \left( \frac{e^{2f+U+V}v^2}{u^2+v^2} +
\frac{e^{2V}u^2}{u^2+v^2}\right) \\
-dx \cdot dv \left( \frac{uy W e^{U+V}}{\sqrt{(x^2+y^2)(u^2+v^2)}} \right)
-dy \cdot du \left( \frac{xv W e^{U+V}}{\sqrt{(x^2+y^2)(u^2+v^2)}}
\right) \\
+dx \cdot du \left( \frac{vy W e^{U+V}}{\sqrt{(x^2+y^2)(u^2+v^2)}} \right)
+dy \cdot dv \left( \frac{ux W e^{U+V}}{\sqrt{(x^2+y^2)(u^2+v^2)}}
\right) \, .
\end{split}
\end{align}
Regularity (avoidance of conical singularities) requires the cross terms 
to vanish on the axes of the Clifford tori, i.e.
\begin{eqnarray} \label{regula}
2f-U+V &= 0 \textrm { \ \ for \ \ } \omega \rightarrow 0 \nonumber \,
, \\ 
2f+U-V &= 0 \textrm { \ \ for \ \ } \omega \rightarrow \frac{\pi}{2}
\, ,
\end{eqnarray}
and
\begin{equation}
W \rightarrow 0 \textrm { \ \ for \ \ } \omega \rightarrow 0 \textrm {
  \ and \ } \omega \rightarrow \frac{\pi}{2} \, .
\end{equation}
The functions $f$ and $W$ should approach zero for $\rho \rightarrow
\infty$ like $f \sim \frac{1}{\rho^{2+\epsilon}}$, and $W \sim
\frac{1}{\rho^{2+\delta}}$ with $\epsilon, \delta > 0$ in order not to
contribute to the ADM mass. (Again, taking derivatives should lower the order
of $\rho$ by one.) The latter is encoded in the lowest order 
term of $U$ and $V$:
\begin{equation} \label{UVasymp}
U \sim \frac{\mu}{\rho^2} = \frac{2M}{3\pi \rho^2} \textrm {\ \ \  and
  \ \ \ } V \sim \frac{\mu}{\rho^2} = \frac{2M}{3\pi \rho^2}
\end{equation}
for $\rho \rightarrow \infty$, where we have defined $\mu =
\frac{2M}{3\pi}$ to keep the notation tidy.

\subsection{Proof of positive mass: Regular data, $W$=0}
In this subsection we perform the proof of positive mass for the case
$W=0$, i.e.~no cross-term in $\alpha$ and $\beta$. Geometrically this
corresponds to an additional $\mathbb{Z}_2$ symmetry. One can
think of the Clifford tori as being rectangular in this case. The data
are supposed to be regular: They have topology $\mathbb{R}^4$ and 
do not contain apparent horizons. \\

We can perform two Kaluza-Klein reductions of the metric (\ref{4met}),
first along $\frac{\partial}{\partial \alpha}$, then along
$\frac{\partial}{\partial \beta}$. The result is
\begin{equation} \label{curveq}
\phantom{}^{(4)}R = \phantom{}^{(2)}R-\frac{2}{X} \left(\nabla_q^2 X
\right)-\frac{2}{Y} \left(\nabla_q^2 Y \right)-\frac{2}{XY} \nabla_q X
\nabla_q Y \, .
\end{equation}
Note that
\begin{align}
\phantom{}^{(2)}R &= -2 e^{-2f-U-V} \bigtriangleup^{(2)}_{flat}
\left(f+\frac{U}{2}+\frac{V}{2}\right) \, , \\
\bigtriangleup^{(2)}_q &= e^{-2f-U-V} \bigtriangleup^{(2)}_{flat} \, , \\
\bigtriangleup^{(4)}_{flat} &= \frac{\partial^2}{\partial \rho^2} +
\frac{3}{\rho}\frac{\partial}{\partial \rho} + \frac{1}{\rho^2}
\frac{\partial^2}{\partial \omega^2}+ \frac{1}{\rho^2}
\left(\frac{\cos \omega}{\sin \omega} -\frac{\sin \omega}{\cos
  \omega}\right)\frac{\partial}{\partial \omega}  \, .
\end{align}
We compute the terms on the right hand side of (\ref{curveq}). We will
omit the index ``flat'' from now on since all gradients and laplacians
refer to the flat metric.
\begin{align}
\frac{\bigtriangleup^{(2)}X}{X} &= \bigtriangleup^{(4)} V + \left(V_{,\rho}\right)^2
+ \frac{V_{,\omega}}{\rho^2} - \frac{V_{,\omega}}{\rho^2}\left(\frac{\cos \omega}{\sin \omega} +\frac{\sin \omega}{\cos
  \omega}\right) \, , \\
\frac{\bigtriangleup^{(2)}Y}{Y} &= \bigtriangleup^{(4)} U + \left(U_{,\rho}\right)^2
+ \frac{U_{,\omega}}{\rho^2} + \frac{U_{,\omega}}{\rho^2}\left(\frac{\cos \omega}{\sin \omega} +\frac{\sin \omega}{\cos
  \omega}\right) \, , \\
\frac{\nabla X \nabla Y}{XY} &= \frac{U_{,\rho} + V_{,\rho}}{\rho}
+ U_{,\rho}V_{,\rho} +
\frac{U_{,\omega}V_{,\omega}}{\rho^2} 
+ \frac{1}{\rho^2}\left(\frac{\cos \omega}{\sin \omega} V_{,\omega}
-\frac{\sin \omega}{\cos \omega} U_{,\omega}\right)  \, , \\
\phantom{}^{(2)}R & \left(-\frac{1}{2}e^{2f+U+V}\right) =
\bigtriangleup^{(2)} \left(f+\frac{U}{2}+\frac{V}{2}\right)
\nonumber \\
&=\Bigg[\bigtriangleup^{(4)} -\frac{2}{\rho}\frac{\partial}{\partial \rho}- \frac{1}{\rho^2}
\left(\frac{\cos \omega}{\sin \omega} -\frac{\sin \omega}{\cos
  \omega}\right)\frac{\partial}{\partial \omega}
\Bigg]\left(f+\frac{U}{2}+\frac{V}{2}\right) \, .
\end{align}
Hence equation (\ref{curveq}) becomes
\begin{equation} \label{curvc}
\begin{split}
-\frac{1}{2} \phantom{}^{(4)}R \cdot e^{2f+U+V} &= \frac{3}{2}\bigtriangleup U +
 \frac{3}{2} \bigtriangleup V + \bigtriangleup f \\
&+\frac{1}{2} \left(\nabla U\right)^2 +\frac{1}{2} \left(\nabla
 V\right)^2+\frac{1}{2} \left(\nabla U+ \nabla V\right)^2 \\
&+\frac{1}{2 \rho^2 \sin \omega \cos \omega} \partial_{\omega}
 \left(U-V\right) \\
&-\frac{2}{\rho} \partial_{\rho} f + \frac{1}{\rho^2}\left(\frac{\sin
 \omega}{\cos \omega} - \frac{\cos
 \omega}{\sin \omega}\right) \partial_{\omega} f \, .
\end{split}
\end{equation}
We now integrate over $\mathbb{R}^4$. The term on the left will be
manifestly negative since the Ricci-scalar is positive by the
constraint equations for maximal data. 
The first two terms on the right hand side
will give $-8 M \pi$ by the assumption (\ref{UVasymp}). The term
$\bigtriangleup f$ vanishes upon integration because 
$\partial_\rho f$ decays like
$\frac{1}{\rho^{3+\epsilon}}$. The terms in the second line are
manifestly positive. Finally, we will now show that the terms in the
last two lines cancel when integrated over $\mathbb{R}^4$. We have
\begin{equation} \label{eins}
4\pi^2 \cdot \frac{1}{2} \int \rho \partial_{\omega} \left(U-V\right) d\rho
d\omega = 4\pi^2 \cdot \frac{1}{2} \int \rho \Big[\left(U-V\right)\left(\rho,
  \frac{\pi}{2}\right)-\left(U-V\right)\left(\rho,0\right)\Big] d\rho
\end{equation}
and 
\begin{equation} \label{zwei}
\begin{split}
- 4\pi^2 \cdot 2 \int \rho^2 f_{,\rho} \sin \omega \cos \omega  d\rho
  d\omega = - 4\pi^2 \cdot 2 \int \Big[\rho^2 f(\rho,\omega)\Big]\Bigg|^{\rho=\infty}_{\rho=0} \sin \omega \cos
  \omega d\omega \\ 
+ 4\pi^2 \cdot 4 \int \rho f \sin \omega \cos \omega d\rho d\omega  \, , 
\end{split}
\end{equation}
and
\begin{equation} \label{drei}
\begin{split}
4 \pi^2 \int \rho \left(\sin^2 \omega - \cos^2 \omega \right)
\partial_{\omega} f d\rho d\omega = \\  4\pi^2 \int \rho \Big[f\left(\rho,
  \frac{\pi}{2}\right) + f\left(\rho,0\right)\Big] d\rho - 4\pi^2
\cdot 4 \int \rho f \sin \omega \cos \omega d\rho d\omega \, .
\end{split}
\end{equation}
The last term of (\ref{drei}) cancels with the last term of
(\ref{zwei}). The sum of the expression (\ref{eins}) and the first
term of (\ref{drei}) vanishes due to the regularity conditions
(\ref{regula}) imposed on the axes. Finally, the first term of
(\ref{zwei}) vanishes because $f$ falls of faster than
$\frac{1}{\rho^2}$ at infinity as assumed above. \\
We can summarise the result in the formula
\begin{equation}
\begin{split}
M = &\frac{1}{16\pi} \int_{\mathbb{R}^4} \phantom{}^{(4)}R \cdot e^{2f+U+V} d^4x + \\
 &\frac{1}{8\pi} \int_{\mathbb{R}^4} \Big[\frac{1}{2} \left(\nabla U\right)^2 +\frac{1}{2} \left(\nabla
 V\right)^2+\frac{1}{2} \left(\nabla U+ \nabla V\right)^2 \Big] d^4x
 \, ,
\end{split}
\end{equation}
which clearly shows $M \geq 0$. 

\subsection{The case $W \neq 0$}
In the $W \neq 0$ case the calculations are much more involved but
straightforward. The analog of (\ref{curvc}) reads

\begin{equation} \label{Brilleq}
\begin{split}
-\frac{1}{2} &\phantom{}^{(4)}R \cdot e^{2f+U+V} = \frac{3}{2}\bigtriangleup U+
 \frac{3}{2} \bigtriangleup V + \bigtriangleup f \\
&+\frac{1}{2} \left(\nabla U\right)^2 +\frac{1}{2} \left(\nabla
 V\right)^2+\frac{1}{2} \left(\nabla U+ \nabla V\right)^2 \\
&+\frac{1}{2 \rho^2 \sin \omega \cos \omega} \partial_{\omega}
 \left(U-V\right) \\
&-\frac{2}{\rho} \partial_{\rho} f + \frac{1}{\rho^2}\left(\frac{\sin
 \omega}{\cos \omega} - \frac{\cos
 \omega}{\sin \omega}\right) \partial_{\omega} f \\
&+ \frac{-W \bigtriangleup W}{1-W^2} - \frac{3}{4} \frac{(\nabla
  W)^2}{1-W^2} - \frac{W^2(\nabla W)^2}{(1-W^2)^2}-\frac{W}{1-W^2}
\frac{1}{\rho}\frac{\partial W}{\partial \rho} \\
&+\frac{1}{4} \frac{W^2}{1-W^2} \Bigg( \left(\nabla U \right)^2 +
 \left(\nabla V \right)^2 - 2 \nabla U
 \nabla V -\frac{6}{W}\left(\nabla U \nabla W + \nabla V \nabla W \right)\Bigg) \\
&+\frac{1}{2 \rho^2} \frac{W^2}{1-W^2}
 \left(\frac{U_{,\omega}}{\sin \omega \cos \omega} -
 \frac{V_{,\omega}}{\sin \omega \cos \omega} - \frac{W_{,\omega}}{W}
 \frac{\cos(2\omega)}{\sin \omega \cos \omega} + \frac{1}{2}
 \frac{1}{\sin^2 \omega \cos^2 \omega} \right) \, .
\end{split}
\end{equation}
We already know how to deal with the terms of the first four lines 
from the previous section. We will now use partial integration to
handle the terms in the fifth line including the function $W$ only. 
We observe that
\begin{equation} \label{helte}
\begin{split}
\int \left(\frac{-W \bigtriangleup W}{1-W^2} - \frac{3}{4} \frac{(\nabla
  W)^2}{1-W^2} - \frac{W^2(\nabla W)^2}{(1-W^2)^2}-\frac{W}{1-W^2}
\frac{1}{\rho}\frac{\partial W}{\partial \rho}\right) \rho^3
 \sin \omega \cos \omega d\rho d\omega d\alpha d \beta = \\
\int \left( \frac{1}{4}\frac{(\nabla W)^2}{1-W^2} + \frac{W^2(\nabla
  W)^2}{(1-W^2)^2} - \log \left(1-W^2\right) \frac{1}{\rho^2} \right)\rho^3
 \sin \omega \cos \omega d\rho d\omega d\alpha d \beta \, .
\end{split}
\end{equation}
All terms on the right 
are manifestly non-negative. We will need the first and the
second term of (\ref{helte}) in the following to make other
expressions manifestly non-negative. Next 
we collect the terms of (\ref{Brilleq}) involving a 
derivative of $U$ or $V$ in $\rho$, borrowing also one of the manifestly 
non-negative terms we just obtained in (\ref{helte}). We will show
non-negativity of the following expression:
\begin{equation} \label{rhoeq}
\begin{split}
\left(U_{,\rho}\right)^2 + \left(V_{,\rho}\right)^2 +
 U_{,\rho}V_{,\rho} + \frac{W^2 \cdot
  \left(W_{,\rho}\right)^2}{(1-W^2)^2} \\
-\frac{3}{2}\frac{W}{1-W^2}\left(U_{,\rho}W_{,\rho}+V_{,\rho}W_{,\rho}\right)
+\frac{1}{4} \frac{W^2}{1-W^2}\left(U_{,\rho}-V_{,\rho}\right)^2 \geq
 0 \, .
\end{split}
\end{equation}
(The last term in the first line of (\ref{rhoeq}) comes from (\ref{helte}).)
Defining $a=U_{,\rho}$, $b=V_{,\rho}$, $c=\frac{W \cdot
  W_{,\rho}}{1-W^2}$ we need to show
\begin{equation}
a^2+b^2+c^2+ab- \frac{3}{2}bc - \frac{3}{2} ac \geq 0 \, .
\end{equation} 
For this we simply write the left hand side as
\begin{equation}
\frac{1}{4}\left((a-b)^2 + c^2\right) +
\frac{3}{4}\left(a+b-c\right)^2 \geq 0 \, .
\end{equation}
Next we turn to the terms of (\ref{Brilleq}) 
involving an $\omega$ derivative and the
term just involving trigonometric functions. Collecting these terms
(again borrowing two manifestly negative terms of (\ref{helte})) we have
to show non-negativity of the expression
\begin{equation} \label{xpr}
\begin{split}
\frac{1}{4 \rho^2} \frac{W^2}{1-W^2}
\left(U_{,\omega}-V_{,\omega}-\frac{W_{,\omega}}{W}\cos(2\omega) +
\frac{1}{\sin \omega \cos \omega}\right)^2  \\
+\Big[-\frac{1}{4\rho^2}\frac{\left(W_{,\omega}\right)^2
  \cos^2(2\omega)}{1-W^2}+\frac{1}{4\rho^2}\frac{\left(W_{,\omega}\right)^2}{1-W^2}\Big]
 \\
+\frac{1}{2}\frac{W U_{,\omega} W_{,\omega}}{\rho^2 (1-W^2)} \cos (2 \omega)
-\frac{1}{2}\frac{W V_{,\omega} W_{,\omega}}{\rho^2 (1-W^2)} \cos (2
\omega) + \frac{W^2 \cdot \left(W_{,\omega}\right)^2}{\rho^2(1-W^2)^2} \\
+\frac{1}{\rho^2}\left(\left(U_{,\omega}\right)^2 +
\left(V_{,\omega}\right)^2 +
U_{,\omega}V_{,\omega}\right)-\frac{3}{2}\frac{W}{\rho^2(1-W^2)}\left(U_{,\omega}W_{,\omega}+V_{,\omega}W_{,\omega}\right)
\geq 0 \, .
\end{split}
\end{equation}
(Note that the second term in the second line and the last term in
the third line of (\ref{xpr}) are taken 
from (\ref{helte}).) The terms in the first and second line of
the expression (\ref{xpr}) are already
manifestly non-negative. For the remaining terms 
we define $A = \frac{U_{,\omega}}{\rho}$, $B =
\frac{V_{,\omega}}{\rho}$ and  $C = \frac{W \cdot
  W_{,\omega}}{\rho(1-W^2)}$ and show that
\begin{equation}
A^2+B^2+C^2 + AB - AC\left(\frac{3}{2}-\frac{1}{2}\cos(2
\omega)\right)
- BC\left(\frac{3}{2}+\frac{1}{2}\cos(2 \omega)\right) \geq 0 \, .
\end{equation}
To see this, we just write the left hand side as
\begin{equation}
\frac{3}{4}\left(A+B-C\right)^2+\frac{1}{4}\left(A-B+C\cos(2\omega)\right)^2+\frac{1}{4}C^2
\left(1-\cos^2(2\omega)\right) \geq 0 \, ,
\end{equation}
making it manifestly non-negative.
We finally obtain the formula
\begin{align} \label{fmass}
\begin{split}
M &= \frac{1}{16\pi} \int_{\mathbb{R}^4} 
\phantom{}^{(4)}R \cdot e^{2f+U+V} d^4x  \\
 &+\frac{1}{8\pi} \int_{\mathbb{R}^4} \Bigg(-\log(1-W^2) \frac{1}{\rho^2} +
 \frac{1}{4} \frac{\left(W_{,\rho}\right)^2}{1-W^2} + \frac{1}{4}
 \frac{1}{\rho^2}\frac{\left(W_{,\omega}\right)^2 \sin^2 (2\omega)}{1-W^2}\\
 & \ \ \ \ \ \ \ +\frac{1}{4}\left( (a-b)^2 + c^2 \right) +
 \frac{3}{4}\left(a+b-c\right)^2 + \frac{3}{4}\left(A+B-C\right)^2 \\
 & \ \ \ \ \ \ \ +\frac{1}{4}\left(A-B+C\cos(2\omega)\right)^2
 + \frac{1}{4}
 C^2 \left(1-\cos^2(2\omega)\right) \\
& \ \ \ \ \ \ \ +\frac{1}{4} \frac{W^2}{1-W^2} \left(A-B-\frac{W_{,\omega}}{W\rho} \cos(2\omega)
  + \frac{1}{\rho \sin \omega \cos \omega}\right)^2 \\
& \ \ \ \ \ \ \ +\frac{1}{4} \frac{W^2}{1-W^2} \left(a-b\right)^2 \Bigg) d^4x \, .
\end{split}
\end{align}

\section{Generalisation to 4+1 dimensions: Black holes}
In this section we generalise the Brill formula to include
black holes and derive a Riemann-Penrose inequality for the
initial data (cf. \emph{Lemma 5.1}). We restrict to the 
case of rectangular tori.

\subsection{The generalised Weyl-Class}
The analog of the four-dimensional Weyl-class for orthogonal tori 
was obtained in \cite{Reall}. Here we will use a slightly 
different representation and state the relevant formulae in both
toroidal and standard $r-z$-orbit space coordinates. The metric is 
\begin{equation} \label{Weyl5}
\begin{split}
g = -e^{-2U\left(\rho,\omega\right)-2V\left(\rho,\omega\right)} dt^2 +
e^{2f(\rho,\omega)+U(\rho,\omega)+V(\rho,\omega)} \left(d\rho^2 +
\rho^2 d \omega^2 \right) + \\ e^{2U(\rho,\omega)}\rho^2
\sin^2 \omega d\alpha^2 + e^{2V(\rho,\omega)}\rho^2 \cos^2 \omega
d\beta^2 \, ,
\end{split}
\end{equation}
with Einstein equations
\begin{eqnarray} \label{Ueq}
\bigtriangleup^{(4)}_{flat} U &= \Big[\frac{\partial^2}{\partial \rho^2} +
\frac{3}{\rho}\frac{\partial}{\partial \rho} + \frac{1}{\rho^2}
\frac{\partial^2}{\partial \omega^2}+ \frac{1}{\rho^2}
\left(\frac{\cos \omega}{\sin \omega} -\frac{\sin \omega}{\cos
  \omega}\right)\frac{\partial}{\partial \omega}\Big] U(r,\omega) = 0
\, ,
\end{eqnarray} 
\begin{eqnarray} \label{Veq}
\bigtriangleup^{(4)}_{flat} V &= \Big[\frac{\partial^2}{\partial \rho^2} +
\frac{3}{\rho}\frac{\partial}{\partial \rho} + \frac{1}{\rho^2}
\frac{\partial^2}{\partial \omega^2}+ \frac{1}{\rho^2}
\left(\frac{\cos \omega}{\sin \omega} -\frac{\sin \omega}{\cos
  \omega}\right)\frac{\partial}{\partial \omega}\Big] U(r,\omega) = 0
\, ,
\end{eqnarray}
and
\begin{equation} \label{feq1}
\begin{split}
-\frac{4}{\rho} \partial_{\rho} f + \frac{4}{\rho^2}
\frac{\cos(2\omega)}{\sin(2 \omega)} \partial_\omega f = &-2
\Big[(\partial_\rho U)^2 + \partial_\rho U  \partial_\rho V +
  (\partial_\rho V)^2 \Big] \\ 
&+ \frac{2}{\rho^2} \Big[ (\partial_\omega U)^2 
+ \partial_\omega U  \partial_\omega V + (\partial_\rho V)^2\Big]
 \\
&+ \frac{1}{\rho^2 \cos \omega \sin \omega} \left(\partial_\omega U -
\partial_\omega V \right) \, ,
\end{split}
\end{equation}
\begin{equation} \label{feq2}
\begin{split}
-\frac{2}{\rho} \frac{\cos(2\omega)}{\sin(2 \omega)} \partial_\rho f -
\frac{2}{\rho^2} \partial_{\omega} f = 
&-\frac{1}{\rho}\left(2 \partial_\rho U \partial_\omega U + 2
\partial_\rho V \partial_\omega V + \partial_\rho V \partial_\omega U
+ \partial_\rho U \partial_\omega V \right) \\
&+ \frac{1}{\rho \sin(2\omega)}
\left(\partial_\rho V - \partial_\rho U \right) \, .
\end{split}
\end{equation}
The relation of the toroidal variables $\rho$ and $\omega$ to the
the standard $r$-$z$ orbit-space variables used in \cite{Reall} is given by
\begin{eqnarray} \label{torico}
r &=& \frac{1}{2} \rho^2 \sin \left(2\omega\right) \\
z &=& \frac{1}{2} \rho^2 \cos \left(2\omega\right)
\end{eqnarray}
such that in $z-r$-coordinates the metric (\ref{Weyl5}) reads
\begin{equation} \label{rzcoord}
\begin{split}
g = -e^{-2U\left(r,z\right)-2V\left(r,z \right)} dt^2 +
e^{2f\left(r,z \right)+U\left(r,z \right)+V\left(r,z\right)-\log
  \left(2\sqrt{r^2+z^2}\right)} \left(dz^2 + dr^2 \right) \\ 
+ e^{2U\left(r,z \right)} \left(-z +
\sqrt{r^2+z^2}\right) d\alpha^2 + e^{2V\left(r,z \right)} \left(z +
\sqrt{r^2+z^2}\right) d\beta^2 \, .
\end{split}
\end{equation}
The equations for the functions $U$ and $V$ translate to
\begin{eqnarray}
2 \sqrt{r^2+z^2} \left[ \partial_r^2 +\frac{1}{r}\partial_r +
  \partial_z^2 \right] U(r,z) = 0 \\
2 \sqrt{r^2+z^2} \left[ \partial_r^2 +\frac{1}{r}\partial_r +
  \partial_z^2 \right] V(r,z) = 0 
\end{eqnarray}
involving the three-dimensional flat Laplacian just like in the
three-dimensional case. It is clear how to translate the equations 
(\ref{feq1}) and (\ref{feq2}) into $r-z$-coordinates. \\

Using equations (\ref{Ueq}), (\ref{Veq}), (\ref{feq1}), (\ref{feq2}) 
and the fact that the Ricci-scalar of the initial data slice 
vanishes because of the time-symmetry of the slice and vacuum, 
one easily checks that the differential Brill-formula (\ref{curvc}) 
holds for the spatial slices in the five-dimensional 
Weyl-class (\ref{Weyl5}).

\subsection{Schwarzschild-Tangherlini initial data}
A conformally flat time-symmetric initial data slice for 
the Schwarzschild-Tangherlini \cite{Tangherlini} metric 
can be obtained by choosing $f=W=0$ and
\begin{equation} \label{confflat}
U = V = \log \left(1+\frac{\mu}{\rho^2} \right) = 
\log \left(1+\frac{2M}{3 \pi \rho^2}  \right)
\end{equation}
in the ansatz (\ref{4met}). This would lead to a spatial
 slice of the form
\begin{equation}
g = \left(1+\frac{2M}{3 \pi \rho^2}\right)^2 \left(d\rho^2 + \rho^2
d\omega_3^2\right) \, ,
\end{equation}
for which we find a minimal surface at $\rho=\sqrt{\frac{2M}{3\pi}}$ 
with area
\begin{equation} \label{confarea}
A = 16 \sqrt{\pi} \left(\frac{2M}{3}\right)^{\frac{3}{2}} \, .
\end{equation}
In particular, this slice covers both asymptotically flat ends. \\
However, this slice is not part of the slicing
implicit in the Weyl-ansatz (\ref{Weyl5}). The Schwarzschild metric in
Weyl-coordinates is given by a certain rod-representation, just as in
the four-dimensional case. We will now turn to this representation,
giving all formulae in $r-z$ coordinates, i.e.~the metric functions
for the ansatz (\ref{rzcoord}). They can easily be translated to toroidal 
coordinates using (\ref{torico}). The harmonic functions $U$ and $V$ are
\begin{equation}
U_S = \frac{1}{2} \log
\frac{\mu-z+\sqrt{\left(\mu-z\right)^2+r^2}}{-z+\sqrt{r^2+z^2}} \, ,
\end{equation}
\begin{equation}
V_S = \frac{1}{2} \log
\frac{\mu+z+\sqrt{\left(\mu+z\right)^2+r^2}}{z+\sqrt{r^2+z^2}} \, .
\end{equation}
The function $U_S$ corresponds to the Newtonian potential induced by a
rod lying on the $z$-axis in the interval $\left[0,\mu\right]$ 
whereas $V_S$ corresponds to a rod in
$\left[-\mu,0\right]$. Asymptotically,
\begin{equation}
U_S \sim \frac{\mu}{2 \sqrt{r^2+z^2}} = \frac{\mu}{\rho^2} 
\textrm{ \ \ \ \ \ \ \ \ } 
V_S \sim \frac{\mu}{2 \sqrt{r^2+z^2}} = \frac{\mu}{\rho^2} \, ,
\end{equation}
as it should be. The function $f_S$ is found to be given by the expression
\begin{equation}
f_S = -\frac{U_S}{2} - \frac{V_S}{2} +
\frac{1}{2}\log\left(\sqrt{r^2+z^2}\right) +\frac{1}{2}
\log\left(\frac{\frac{r_1+r_2}{2}+\mu}{r_1r_2}\right) \, ,
\end{equation}
where we have defined the distances from the rod
\begin{equation}
r_1 = \sqrt{\left(z+\mu\right)^2+r^2} \textrm{\ \ \ \ \ \ \ \ \ \ } 
r_2 = \sqrt{\left(z-\mu\right)^2+r^2}
\end{equation}
analogously to the four-dimensional case. The function $f_S$ is unique
up to a constant, which is determined by asymptotic flatness. 
Asymptotically,
\begin{equation}
f_S \sim \frac{\mu^2}{2\left(r^2+z^2\right)} = \frac{2\mu^2}{\rho^4}
\, ,
\end{equation}
as required. Inserting $U_S$, $V_S$ and $f_S$ into (\ref{rzcoord}) 
will yield the Schwarzschild
 metric. Transforming to toroidal coordinates (\ref{torico}) 
and then applying the coordinate transformation (cf. \cite{Reall})
\begin{eqnarray}
\omega &=& \frac{1}{2} \arctan
\left[\frac{\sqrt{1-\frac{4\mu}{R^2}}}{1-\frac{2\mu}{2R^2}} \tan
  \left(2\theta\right) \right] \\
\rho &=& \left(R^4 - 4\mu R^2 + 4\mu^2 \cos \left(2 \theta
\right)\right)^{\frac{1}{4}}
\end{eqnarray}
will finally lead to the Schwarzschild metric in the standard 
$(t,R, \theta, \alpha, \beta)$ coordinates:
\begin{equation}
g_S = -\left(1-\frac{4\mu}{R^2}\right)dt^2 +
\left(1-\frac{4\mu}{R^2}\right)^{-1} dR^2 + R^2 \left(d\theta^2 +
\sin^2 \theta d\alpha^2 + \cos^2 \theta d\beta^2 \right)
\end{equation}
where $\mu=\frac{2M}{3\pi}$ as defined above. \\
As a check we compute the area of the rod using (\ref{Weyl5}). The right part 
of the rod, located on the $z$-axis between $0$ and $\mu$, has area
\begin{equation} \label{areatang}
A = \int_0^{\sqrt{2\mu}} \left(e^{f_S+\frac{3}{2}U_S+\frac{3}{2}V_S}
\rho^2 \cos \omega \sin \omega \right) d\rho d\alpha d\beta \, .
\end{equation}
Note that this will only give \emph{half} of the area of the full
rod, since we also need to take the left part (located at
$\omega=\frac{\pi}{2}$) into account. 
Inserting the functions $U_S$, $V_S$ and $f_S$ defined 
above, we find that
\begin{equation}
e^{f_S+\frac{3}{2}U_S+\frac{3}{2}V_S} \rho^2 \cos \omega \sin \omega 
\Big|_{rod} = 2 \sqrt{\mu} \rho 
\end{equation}
and hence
\begin{equation}
A = 8 \sqrt{\pi} \left(\frac{2M}{3}\right)^{\frac{3}{2}} \,
\end{equation}
which -- if we take \emph{both} halves of the rod into account -- 
is consistent with (\ref{confarea}), the area obtained for the 
conformally flat slice. \\ \\
Finally, using (\ref{areatang}) one checks that the rod 
between $-\mu$ and $\mu$ is indeed a minimal surface, namely that
\begin{equation} 
\frac{1}{\rho} \frac{\partial}{\partial \omega} A = 
\pm \rho \frac{\partial}{\partial r} A = 0 \, 
\end{equation}
holds on the rod.

\subsection{Brill-formula}
Just as in the four-dimensional case, we can consider deformations of
the Schwarzschild initial data:
\begin{eqnarray}
U = U_S + \tilde{U} \textrm { \ \ \ \ \ \ \ }
V = V_S + \tilde{V} \textrm { \ \ \ \ \ \ \ }
f = f_S + \tilde{f} \, ,
\end{eqnarray}
where we assume the following asymptotics near infinity
\begin{equation}
\tilde{U} \sim \frac{2\Delta M}{3 \pi \rho^2} 
\textrm{\ \ \ \ \ \ \ \ \ } 
\tilde{V} \sim \frac{2\Delta M}{3 \pi \rho^2} 
\textrm{ \ \ \ \ \ \ \ \ \ } 
\tilde{f} \sim \frac{1}{\rho^{2+\epsilon}}  \, .
\end{equation}
and that taking derivatives lowers the order of $\rho$ by
 $1$. Integrating the differential Brill-formula 
(\ref{curvc}) will now lead to 
\begin{equation} \label{Brill4}
\begin{split}
\Delta M &= \frac{1}{16 \pi} \int \phantom{}^{(4)} R e^{2f+U+V} +
\frac{1}{8\pi} \int \Big[\frac{1}{2} \left(\nabla \tilde{U}
  \right)^2 + \frac{1}{2} \left(\nabla \tilde{V} \right)^2 \frac{1}{2}
  \left(\nabla \tilde{U} + \nabla \tilde{V} \right)^2 \Big] \\
&+\frac{1}{2} \pi \int_0^{\sqrt{2 \mu}} \Big[\rho \sin \omega \cos \omega
  \partial_\omega \left(\frac{3}{2} \tilde{U} +
\frac{3}{2} \tilde{V} + \tilde{f} \right) \Big]
\left(\rho, \frac{\pi}{2} \right) d\rho \\
&-\frac{1}{2}\pi \int_0^{\sqrt{2 \mu}} \Big[\rho \sin \omega \cos \omega
  \partial_\omega \left(\frac{3}{2} \tilde{U} +
\frac{3}{2} \tilde{V} + \tilde{f} \right) \Big]
\left(\rho, 0 \right) d\rho \\
&+\frac{1}{2}\pi \int_0^{\sqrt{2 \mu}} \rho
\left(\frac{3}{2} \tilde{U} +\frac{3}{2} \tilde{V} + \tilde{f}
\right)\left(\rho, \frac{\pi}{2} \right) d\rho \\
&+\frac{1}{2}\pi \int_0^{\sqrt{2 \mu}} \rho
\left(\frac{3}{2} \tilde{U} +\frac{3}{2} \tilde{V} + \tilde{f}
\right)\left(\rho, 0 \right) d\rho \, .
\end{split}
\end{equation}
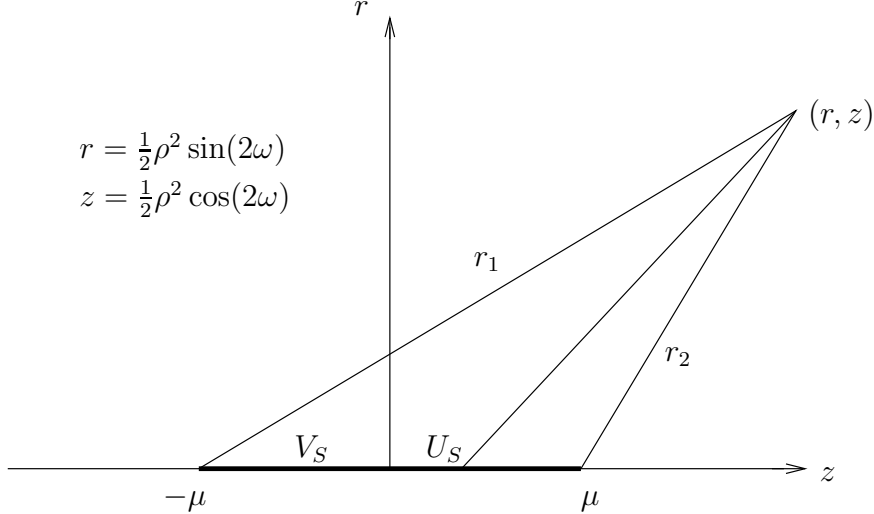
\[
\begin{picture}(0,0)%
\includegraphics{Weyl4.pstex}%
\end{picture}%
\setlength{\unitlength}{3947sp}%
\begingroup\makeatletter\ifx\SetFigFont\undefined%
\gdef\SetFigFont#1#2#3#4#5{%
  \reset@font\fontsize{#1}{#2pt}%
  \fontfamily{#3}\fontseries{#4}\fontshape{#5}%
  \selectfont}%
\fi\endgroup%
\begin{picture}(5575,3256)(1189,-3035)
\put(6301,-2836){\makebox(0,0)[lb]{\smash{{\SetFigFont{12}{14.4}{\rmdefault}{\mddefault}{\updefault}{\color[rgb]{0,0,0}$z$}%
}}}}
\put(4126,-1486){\makebox(0,0)[lb]{\smash{{\SetFigFont{12}{14.4}{\rmdefault}{\mddefault}{\updefault}{\color[rgb]{0,0,0}$r_1$}%
}}}}
\put(5326,-2086){\makebox(0,0)[lb]{\smash{{\SetFigFont{12}{14.4}{\rmdefault}{\mddefault}{\updefault}{\color[rgb]{0,0,0}$r_2$}%
}}}}
\put(1651,-811){\makebox(0,0)[lb]{\smash{{\SetFigFont{12}{14.4}{\rmdefault}{\mddefault}{\updefault}{\color[rgb]{0,0,0}$r = \frac{1}{2}\rho^2 \sin(2\omega)$}%
}}}}
\put(1651,-1111){\makebox(0,0)[lb]{\smash{{\SetFigFont{12}{14.4}{\rmdefault}{\mddefault}{\updefault}{\color[rgb]{0,0,0}$z = \frac{1}{2}\rho^2 \cos(2\omega)$}%
}}}}
\put(3001,-2686){\makebox(0,0)[lb]{\smash{{\SetFigFont{12}{14.4}{\rmdefault}{\mddefault}{\updefault}{\color[rgb]{0,0,0}$V_S$}%
}}}}
\put(3826,-2686){\makebox(0,0)[lb]{\smash{{\SetFigFont{12}{14.4}{\rmdefault}{\mddefault}{\updefault}{\color[rgb]{0,0,0}$U_S$}%
}}}}
\put(3376, 89){\makebox(0,0)[lb]{\smash{{\SetFigFont{12}{14.4}{\rmdefault}{\mddefault}{\updefault}{\color[rgb]{0,0,0}$r$}%
}}}}
\put(2176,-2986){\makebox(0,0)[lb]{\smash{{\SetFigFont{12}{14.4}{\rmdefault}{\mddefault}{\updefault}{\color[rgb]{0,0,0}$-\mu$}%
}}}}
\put(4801,-2986){\makebox(0,0)[lb]{\smash{{\SetFigFont{12}{14.4}{\rmdefault}{\mddefault}{\updefault}{\color[rgb]{0,0,0}$\mu$}%
}}}}
\put(6226,-586){\makebox(0,0)[lb]{\smash{{\SetFigFont{12}{14.4}{\rmdefault}{\mddefault}{\updefault}{\color[rgb]{0,0,0}$(r,z)$}%
}}}}
\end{picture}%

\]
which can be translated to $r-z$ coordinates, where we find 
the expression familiar from the three-dimensional case 
(cf. equation (\ref{deltam}))
\begin{equation} \label{Brill5}
\begin{split}
\Delta M = \frac{1}{16 \pi} &\int_{vol} \phantom{}^{(4)} R e^{2f+U+V} +
\frac{1}{8\pi} \int_{vol} \Big[\frac{1}{2} \left(\nabla \tilde{U}
  \right)^2 + \frac{1}{2} \left(\nabla \tilde{V} \right)^2 \frac{1}{2}
  \left(\nabla \tilde{U} + \nabla \tilde{V} \right)^2 \Big] \\
&-\pi \int_{-\mu}^{\mu} r \frac{\partial}{\partial r} \left(\frac{3}{2} \tilde{U} +\frac{3}{2}
\tilde{V} + \tilde{f} \right) \Bigg|_{r=0} dz \\
&+\pi \int_{-\mu}^{\mu} \left(\frac{3}{2} \tilde{U} +\frac{3}{2}
\tilde{V} + \tilde{f} \right) \Bigg|_{r=0} dz \, .
\end{split}
\end{equation}
To obtain these formulae one uses
\begin{equation}
-\vec{n} \cdot \vec{\nabla}\Big|_{rod} = \mp \frac{1}{\rho} \frac{\partial}{\partial
  \omega} = -\rho \frac{\partial}{\partial r} \, ,
\end{equation}
where the upper (lower) sign stands for right (left) part of the rod, 
$\omega = 0$ ($\omega = \frac{\pi}{2}$). The vector $-\vec{n}$ is the 
unit-normal vector on the rod. Furthermore,
\begin{displaymath}
r \frac{\partial}{\partial r} U_S \Big|_{z-axis} = \left\{ \begin{array}{ll}
-1 & \textrm{on the right part of the rod}  \\
0 & \textrm{else}
\end{array} \right.
\end{displaymath}
\begin{displaymath}
r \frac{\partial}{\partial r} V_S \Big|_{z-axis} = \left\{ \begin{array}{ll}
-1 & \textrm{on the left part of the rod}  \\
0 & \textrm{else}
\end{array} \right.
\end{displaymath} \\ \\
From (\ref{Brill4}) one infers the following analogue of the
\emph{Lemma 3.1}:
\\
\\
\emph {Lemma 5.1}: For $T^2$-symmetric deformations 
of Schwarzschild-Tangherlini initial data 
$(\tilde{f},\tilde{U},\tilde{V})$ admitting support
only outside the minimal surface,
i.e.~$\tilde{U}=\tilde{V}=\tilde{f}=0$ 
on the rod,  the mass of the new initial data
will increase, whilst the area will stay the same. Hence
\begin{equation}
M \geq \frac{3}{2} \left(\frac{A}{16 \sqrt{\pi}}\right)^{\frac{2}{3}}
\end{equation}
holds for deformations away from the minimal surface. 
\\
\\
To prove the four-dimensional Riemannian-Penrose inequality 
in full generality in the $T^2$-symmetric setting it remains 
to show that any biaxisymmetric initial 
data can be written in deformed Schwarzschild form such that
$\frac{3}{2}\tilde{U}+\frac{3}{2}\tilde{V}+\tilde{f}=0$
 always holds on the rod, leading us to the same obstacles encountered
 in section 3.

\section{Conclusion}
We have presented various generalisations of an argument used by Brill
to prove positive mass of three-dimensional, 
axisymmetric, time-symmetric, regular initial-data for Einstein's
equations. We first noted that the symmetry assumptions can be
slightly weakened.  Furthermore an apparent horizon was introduced in
the data, which was seen to be represented by a rod, when the data was
put into Weyl's form. A simple calculation then established a version
of the Riemannian Penrose inequality, namely that all axisymmetric
deformations (not necessarily small) of the Schwarzschild metric away
from the horizon will increase the mass while leaving the area
invariant. We indicated how we hope to proceed with an elementary
proof of the general case in this axisymmetric class. While the
Penrose-inequality has been proven for the time-symmetric case in
great generality, there is so far not much
mathematical evidence that it will hold in higher dimensions. Here we
were able to show the version of the Riemannian
Penrose inequality mentioned above for four-dimensional initial data
admitting an action by the torus group $T^2$. For the proof we derived
the generalisation of Brill's formula for $4+1$ dimensions and also
obtained the positive mass theorem for data with trivial topology. \\ \\
Although the primary objective will be the removal of
the gap preventing us to prove the Riemannian Penrose inequality 
in full generality using our method of rods, the techniques
used in this paper should also allow generalisation to further
interesting cases, which we hope to address in the future. On the one
hand, one could include more than one apparent horizon and try to
extract quantitative information about the interaction energies and
the gravitational radiation produced in the development of the data
(see \cite{Araujo} for an attempt in this direction). On the other
hand, one could enquire about the inclusion of a cosmological constant
or a study of the electrovacuum case. In the latter case, one may hope
to prove the stronger inequality
\begin{equation}
M \geq \sqrt{\frac{A}{16 \pi}} + q^2 \sqrt{\frac{\pi}{A}} \,
\end{equation}
as suggested in \cite{Gibbons3}. All generalisations mentioned can be
attempted for both $3+1$ and $4+1$ dimensions.

\section{Acknowledgement}
We thank Claude Warnick for a careful reading of the
manuscript. G.H. thanks EPSRC and Studienstiftung des deutschen Volkes
for financial support.

\end{document}